\documentclass[pra,twocolumn,showpacs]{revtex4}

\usepackage{graphicx}
\usepackage{hyperref}
\usepackage{amssymb,amsmath,amsfonts}
\usepackage{array}
\usepackage{color}
\usepackage{euscript}

\def\ds{\displaystyle}

\begin{document}

\title{Generation of two-color polarization-entangled
optical beams with a self-phase-locked two-crystal Optical
Parametric Oscillator}

\author{Julien Laurat\dag, Ga{\"e}lle Keller\ddag, Claude
Fabre\ddag, Thomas Coudreau\ddag\S\footnote[3]{To whom
correspondence should be addressed (coudreau@spectro.jussieu.fr)}}

\affiliation{\dag Norman Bridge Laboratory of Physics 12-33,
California Institute of Technology, Pasadena, CA 91125,USA}

\affiliation{\ddag Laboratoire Kastler Brossel, Universit{\'e}
Pierre et Marie Curie, Case 74, 4 Place Jussieu, 75252 Paris cedex
05, France}

\affiliation{\S Laboratoire Mat{\'e}riaux et Ph{\'e}nom{\`e}nes
Quantiques, Universit{\'e} Denis Diderot, Case 7021, 2 Place
Jussieu, 75251 Paris cedex 05, France}

\begin{abstract}
A new device to generate polarization-entangled light in the
continuous variable regime is introduced. It consists of an
Optical Parametric Oscillator with two type-II phase-matched
non-linear crystals orthogonally oriented, associated with
birefringent elements for adjustable linear coupling. We give in
this paper a theoretical study of its classical and quantum
properties. It is shown that two optical beams with adjustable
frequencies and well-defined polarization can be emitted. The
Stokes parameters of the two beams are entangled. The principal
advantage of this setup is the possibility to directly generate
polarization entangled light without the need of mixing four modes
on beam splitters as required in current experimental setups. This
device opens new directions for the study of light-matter
interfaces and generation of multimode non-classical light and
higher dimensional phase space.
\end{abstract}

\date{\today}

 \maketitle

\section{Introduction}

Developing quantum memories constitutes an essential milestone on
the route towards quantum communication networks. As it is
difficult to directly store photons, the quantum information has
to be stored in a material-based quantum system. A particularly
promising application for polarization-entangled light is the
ability to couple it with atomic ensembles: the algebra describing
the quantum properties of polarized light \emph{via} Stokes
operators \cite{chirkin93,korolkova02,korolkova05} is exactly the
same as that describing the quantum properties of atomic spin be
it single or collective spins. Quantum state exchange between
light fields and matter systems has been recently experimentally
demonstrated \cite{memoirepolzik,reviewlukin} and it has been
shown that such systems can be used as quantum memories
\cite{memoiredantan}. In addition, from a technical point of view,
the use of polarization states offers simpler detection schemes,
without the need for local oscillators. These features make the
study and experimental generation of non-classical polarization
states of first importance for continuous variable quantum
communication.

\begin{figure*}[htpb!]
\centerline{\includegraphics[width=.7\columnwidth]{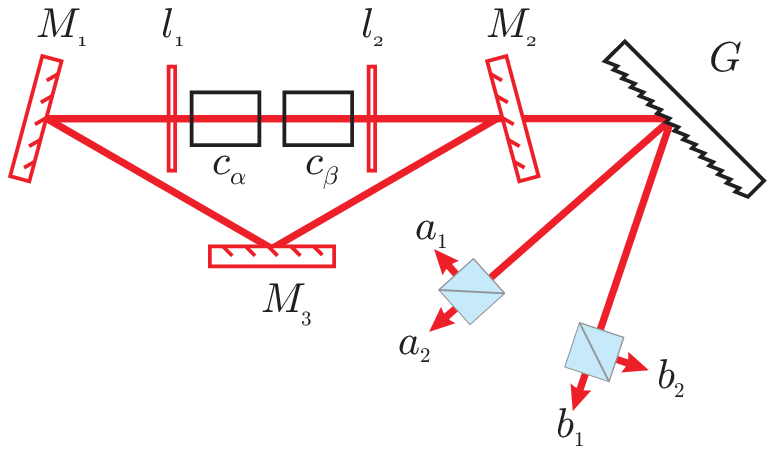}%
\includegraphics[width=.4\columnwidth]{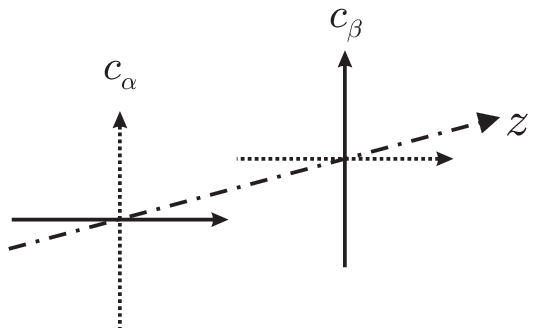}%
\includegraphics[width=.8\columnwidth]{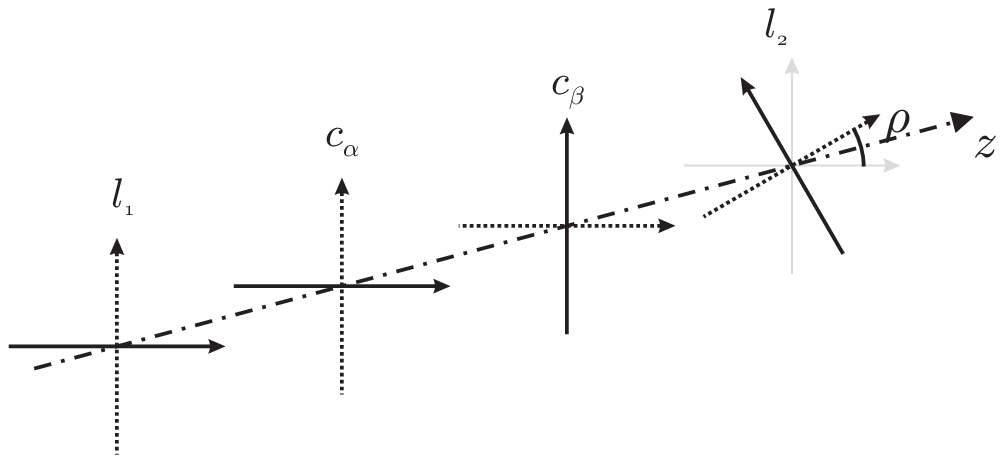}}
\caption{\label{fig:cavite} Left: Ring cavity two-crystal OPO :
$c_{\alpha,\beta}$ are type-II nonlinear crystals, $l_{1,2}$ are
$\lambda/2$ waveplates at $\frac{\omega_0}{2}$ and $\lambda$
waveplates at $\omega_0$. $M_{1,3}$ are highly reflective mirrors
at $\omega_{a,b}$ and $M_2$ is a partially reflective mirror at
$\omega_{a,b}$. All mirrors are transparent at $\omega_0$. $G$ is
a diffraction grating which separates the two frequency
components. Each of them have two orthogonally-polarized bright
components. Center: neutral axes seen by the fundamental wave at
frequency $\omega_0$. Right: neutral axes seen by the sub-harmonic
waves at frequencies $\omega_1$ and $\omega_2$. Dotted lines
denote fast axes and continuous lines slow axes.}
\end{figure*}

Furthermore, the study of higher dimensional phase space is a
fascinating and promising subject. Indeed, it has been shown that
multimode non-classical light could be a very powerful tool for an
efficient processing and transport of quantum information
\cite{gatti04}. Already experimental implementations and
applications of such non-classical multimode light have been
demonstrated in the continuous-wave regime
\cite{martinelli03,treps03,lugiato04,lantz05}. Such systems either
involve a very large number of modes \cite{lugiato04,lantz05} or a
much smaller number (three in the case of ref. \cite{treps03}).
Increasing the number of correlated modes is of particular
importance, for instance in the case of optical read-out
technologies \cite{cd} and super-resolution techniques
\cite{super}. In this context, generating continuous-variable
polarization-entangled light
\cite{chirkin93,korolkova02,korolkova05}, which exhibits three or
four mode correlations, is a first step towards increasing the
phase space dimension.

Such polarization-entangled beams in the continuous variable
regime have been recently produced experimentally with injected
type-I phase-matched Optical Parametric Oscillators (OPOs) below
threshold
\cite{polarentanglement_bowen,polarentanglement_schnabel} or using
$\chi^{(3)}$ effects in a cloud of cold atoms
\cite{polarentanglement_josse}. All these methods require linear
interferences of two quadrature-entangled modes with two bright
coherent states, which limits the amount of entanglement reached
and the simplicity and scalability of the setup.

We investigate here a new approach to directly generate
polarization-entangled light without the need of mode mixing. This
approach stems from an original device we previously studied, a
self-phase-locked optical parametric oscillator. This system
consists of a type-II phase-matched nonlinear $\chi^{(2)}$ crystal
in an optical cavity and emits fields that are phase-locked
through linear coupling \cite{wong98,fabre99,epjdclassique}: a
birefringent plate, which can be rotated relative to the principal
axis of the crystal, adds this coupling between the orthogonally
polarized signal and idler beams and results in a phase-locking
phenomenon that is well-known for coupled mechanical or electrical
oscillators \cite{synchro}. Such systems have recently attracted a
lot of attention as efficient sources of non-classical light
\cite{longchambon02,adamyan04,epjdquantique}. Stable operation has
been demonstrated experimentally even with very small coupling
\cite{wong98,above,hollandais} and their non-classical properties
are very encouraging \cite{above}. We propose here a new
implementation where two crystals, and not only one, are placed
with their neutral axis orthogonal in a cavity containing two
birefringent waveplates for adjustable linear coupling. Each
crystal generates a pair of signal and idler beams and the
waveplates couple together the two signal beams and the two idler
beams so that phase locking can be achieved in a non-frequency
degenerate operation. As we will show below, this device allows
for a direct and efficient generation of polarization-entangled
optical beams.

In this paper, we investigate in detail the classical and quantum
properties of the system. The paper is organized as follows. In
Sec. \ref{sec2}, we begin by presenting the linear and non-linear
elements of the two-crystal OPO. A propagating Jones matrix is
associated to each elements and permits to access the round-trip
matrix and the stationary solutions. Section \ref{sec3} considers
the quantum properties of the emitted beams in terms of quadrature
operators in the Fourier domain. The usual input-output
linearization technique is used and quantum correlation spectra
are formulated. In Section \ref{sec4}, these quadrature
correlations are then interpreted in terms of
polarization-entanglement through the Stokes operators.
Considering various criteria, we demonstrate theoretically that
this device is an efficient source of polarization-entanglement.

\section{Setup and classical properties}
\label{sec2}

In this section, we present the basic scheme of the two-crystal
OPO. Round-trip equations and stationary solutions are then
derived by using Jones matrix formalism.

\subsection{Linear and nonlinear elements\\in the OPO cavity}

The setup is sketched in Fig. \ref{fig:cavite}. Two identical
type-II phase matched ($\chi^{(2)}$) crystals oriented at
90$^{\circ}$ and two birefringent waveplates are inserted inside a
ring cavity. Signal and idler fields are resonant but the pump is
not enhanced. The orientation of the various axes with respect to
one another are critical. They are summarized on Fig.
\ref{fig:cavite}:
\begin{itemize}
\item the nonlinear crystals $c_\alpha$ and $c_\beta$ have their fast
axes orthogonal; \item waveplate $l_1$ has its fast axis parallel
to the fast axis of the crystal $c_\alpha$; \item the fast axis of
the waveplate $l_2$ makes an angle $\rho$ with respect to the fast
axis of crystal $c_\beta$. In this paper, we will restrict
ourselves to small values of $\rho$.
\end{itemize}

Four fields may propagate inside the cavity: their classical field
amplitudes are denoted $a_{1,2}$ and $b_{1,2}$ where the index
corresponds to the polarization and the letter to the frequency
$\omega_a$ or $\omega_b$ (Fig. \ref{fig:cavite}). The pump field
is polarized at 45$^\circ$ of the crystals' axes and its amplitude
is denoted $a_0$. We consider that the pump and subharmonic fields
are coupled via the nonlinear crystals. In $c_\alpha$, the phase
matching is such that $a_0^{(x)}$ is coupled to $a_1$ and $b_2$
while in $c_\beta$, $a_0^{(y)}$ is coupled to $a_2$ and $b_1$.

We suppose that the waveplates for the sub-harmonic waves have
identical dephasing for the two fields. $M_1$ and $M_3$ are highly
reflective mirrors while the amplitude reflection coefficient for
$M_2$ is denoted $r$. For the sake of simplicity, this coefficient
is assumed to be real and independent of the frequency and of the
polarization.

\subsection{Propagation matrices}

Birefringent elements are usually described by Jones matrices
\cite{jones41}. The total field can be decomposed into four
components corresponding to the two sub-harmonic frequencies and
the two linear polarizations. The waveplates $l_1$ and $l_2$
couple the fields $a_1$ and $a_2$, and $b_1^\ast$ and $b_2^\ast$
independently while the crystals couple $a_1$ and $b_2^\ast$, and
$a_2$ and $b_1^\ast$ independently. Thus, the basis $\{a_1,
b_2^\ast, a_2, b_1^\ast\}$ is sufficient.

The waveplate $l_1$ is a $\lambda/2$ with its axes parallel to the
basis axes so that its associated matrix is diagonal:

\begin{equation}
\small
M_{l_1} = \left( \begin{array}{cccc} i e^{i k_a n e} & 0 & 0 & 0\\
0 & -i e^{-i k_b n e} & 0 & 0\\
0 & 0 &i e^{i k_a n e} & 0\\
0 & 0 & 0 & -i e^{-i k_b n e}
\end{array}\right)
\end{equation}
where $k_{a,b}$ denotes the wave vectors, $n$ the mean index of
refraction (dispersion is neglected) and $e$ the thickness of the
plate.

To determine the transfer matrix of the nonlinear crystals, the
phase-matching will be taken perfect. We recall that the field
amplitudes at the output of the first crystal can be written
\begin{eqnarray}\label{eq2}
a_1^{(out)} &=& a_1^{(in)} + g a_0^{(x)}
\left(b_2^{(in)}\right)^\ast \nonumber\\ \mbox{and} \quad
\left(b_2^{(out)}\right)^\ast &=& \left(b_2^{(in)}\right)^\ast + g
\left(a_0^{(x)}\right)^\ast \left(a_1^{(in)}\right)
\end{eqnarray}
to the first order in $g$ where $a_0^{(x)}$ denotes the pump
amplitude at the output of the first crystal,
\begin{eqnarray}
a_0^{(x)} = \frac{1}{\sqrt 2} a_0 \end{eqnarray} and where $g$ is
the nonlinear coupling coefficient given by
\begin{equation}
g = l \chi^{(2)} \sqrt{\frac{\hbar \omega_0 \omega_a \omega_b}{2
c^2 \varepsilon_0 n_0 n_1 n_2}}.
\end{equation}
Equations \ref{eq2} are only valid close to the oscillation
threshold where the pump depletion is small, which corresponds to
the usual operation regime of such non-linear devices in quantum
optics \cite{above}. The transfer matrix of the first crystal
takes thus the following form:
\begin{equation}
\small M_{c_{\alpha}} = \left(\begin{array}{cccc} e^{i k_a n_1 l}
& g \frac
{a_0}{\sqrt 2} e^{ i k_a n_1 l} & 0 & 0 \\
g \frac{a_0^\ast }{\sqrt 2}e^{- i k_b n_2 l} & e^{- i k_b n_2 l} & 0 & 0 \\
0 & 0 & e^{ i k_a n_2 l} & 0 \\
0 & 0 & 0 & e^{- i k_b n_1 l}
\end{array}\right).
\end{equation}
As stated before, perfect phase-matching has been assumed.
$n_{1,2}$ stand for the refractive indices along the neutral axes
and $l$ is the crystal thickness. In the second crystal, only
$a_2$ and $b_1^\ast$ are coupled to the pump. Assuming that the
two crystals are at the same temperature, the transfer matrix is
given by:
\begin{equation}
\small
M_{c_{\beta}} = \left(\begin{array}{cccc} e^{i k_a n_2 l} & 0 & 0 & 0 \\
0 & e^{-i k_b n_1 l} & 0 & 0 \\
0 & 0 & e^{i k_a n_1 l} & g a_0^{(y)} e^{i k_a n_1 l} \\
0 & 0 & g a_0^{(y)} e^{-i k_b n_2 l} &e^{-i k_b n_2 l}
\end{array}\right)
\end{equation}
$a_0^{(y)}$ denotes the pump amplitude along $y$ between the two
crystals, \begin{eqnarray} a_0^{(y)} = \frac{1}{\sqrt 2} a_0.
\end{eqnarray}

Finally, the last waveplate is a $\lambda/2$ waveplate rotated by
an angle $(\rho + \pi/2)$ with respect to $l_1$. To the first
order in $\rho$ , the transfer matrix is found to be
\cite{epjdclassique}
\begin{equation}
\small M_{l_2} = \left(\begin{array}{cccc} -i e^{i k_a n
e} & 0 & i \epsilon_0 e^{i k_a n e} & 0\\
0 &  i e^{- i k_b n e} & 0 & -i
\epsilon_0 e^{-i k_b n e} \\
i \epsilon_0 e^{i k_a n e} & 0 & i
e^{i k_a n e} & 0\\
0 & - i \epsilon_0 e^{- i k_b n e} & 0 & -i e^{i k_b n e}
\end{array}\right)
\end{equation}
where $\epsilon_0 \equiv \sin 2 \rho \approx 2 \rho$.

As mentioned earlier, we assume that the mirror properties are
independent both of the frequency and of the polarization. The
free propagation matrix is thus
\begin{equation}
\small
M_{prop} = r \left(\begin{array}{cccc} e^{i k_a L} & 0 & 0 & 0\\
0& e^{-i k_b L} & 0 & 0 \\ 0 & 0 & e^{i k_a L} & 0\\ 0 & 0 & 0 &
e^{-i k_b L}
\end{array}\right)
\end{equation}
where $L$ is the cavity length without taking into account the
birefringent elements (waveplates and crystal).

One can now calculate the round-trip matrix:
\begin{equation}
M_{rt} \equiv
M_{prop}.M_{l_2}.M_{c_{\beta}}.M_{c_{alpha}}.M_{l_1}.
\end{equation}
Rather than giving the round-trip matrix in the general case, let
us now make the following assumptions which are usually verified
experimentally:
\begin{itemize}
\item the finesse for the subharmonic fields is high. The
coefficient $r$ is close to 1 and we put $r=1- \kappa$ with
$\kappa \ll 1$; \item we assume that the double resonance
condition is verified, that is both
\begin{eqnarray}\Delta _{a,b} = k_{a,b} (L + 2 n e +l
(n_1+n_2))\end{eqnarray} are close to an integer multiple of
$2\pi$. We denote $\delta_{a,b}$ their small round-trip phase
detunings. \item the waveplates $l_1$ and $l_2$ are $\lambda/2$
for the two frequencies $\omega_a$ and $\omega_b$.
\end{itemize}

In this case, the round-trip matrix to the first order in
$\delta_{a,b}, g, \kappa$ and $\epsilon_0$ takes the following
form:
\begin{equation}
\small
M_{rt} \approx \left(\begin{array}{cccc} 1 + i \delta_a-\kappa & g
\frac{a_0}{\sqrt 2} & \epsilon_0 & 0 \\
g \frac{a_0^\ast}{\sqrt 2}  & 1 - i \delta_b - \kappa & 0 & -
\epsilon_0\\
- \epsilon_0 & 0 & 1 + i \delta_a-\kappa & g \frac{a_0}{\sqrt 2}
e^{-i
\psi} \\
0 & \epsilon_0 & g \frac{a_0^\ast}{\sqrt 2} e^{i\psi}& 1-i
\delta_b- \kappa
\end{array}\right)
\end{equation}
where $\psi = (k_b n_1 + k_a n_2) l$. The phase of the pump
amplitude between the two crystals has been shifted by $ (k_a+k_b)
n e$.

\subsection{Stationary solutions}

The stationarity of the solutions corresponds to the fact that the
field amplitudes after one round trip must be equal to the initial
ones. This condition can be expressed in terms of the round-trip
matrix:
\begin{equation}
M_{rt} \left(\begin{array}{c} a_1\\b_2^\ast\\ a_2\\b_1^\ast
\end{array}\right) = \left(\begin{array}{c} a_1\\b_2^\ast\\
a_2\\b_1^\ast \end{array}\right)
\end{equation}
This system has a non-trivial solution only if
\begin{equation}
\det (M_{rt}-\mathbb{I}_4) = 0 \label{eq:det=0}
\end{equation}
which leads to a stringent condition on the dephasings
\begin{eqnarray}
\delta_a=\delta_b=\delta
\end{eqnarray}
and provides also an expression for the pump power
$I_0^{(threshold)}=|a_0|^2$.

We will restrict ourselves to the case where $\psi=0[2\pi]$, which
corresponds to the lowest threshold area. Eq \ref{eq:det=0} yields
to:
\begin{eqnarray}
I_0^{(threshold)}&=&|a_0|^2=\frac{2}{g^2}\left(\kappa^2+(\delta\pm\epsilon_0)^2\right).\label{condstat}
\end{eqnarray}
This equation has two solutions, corresponding to two possible
regimes. In this paper, we will focus on the regime of lower
threshold.

These equations are similar to that obtained in a one-crystal
self-phase locked OPO, which has been studied in detail in
\cite{fabre99,epjdclassique,hollandais}. For a given input pump
intensity, the phase-locked operation can be obtained only for a
given range of the relevant parameters (\emph{i.e} cavity length
and crystal temperature) which defines a so-called "locking zone".
This locking zone is typically a cavity resonance width large in
terms of length and a few Kelvins in terms of temperature, well
within experimental capacities. For a one-crystal self-phase
locked OPO, phase-locked operation was indeed observed even for
extremely small angles of the wave plate \cite{above} where it
does not perturb the quantum properties of the system
\cite{epjdquantique}.

Let us consider the lowest threshold given by Eq. \ref{condstat}
and reached for $\delta=\epsilon_0$. This working point,
associated with $\psi=0[2\pi]$, can be obtained experimentally by
simultaneously adjusting both the crystal temperatures and the
frequency of the pump laser. Equation \ref{condstat} becomes thus:
\begin{equation}
I_0^{(threshold)} = 2 \kappa^2/g^2. \label{eq:solstatpompe}
\end{equation} At this working point, the round-trip matrix is
simpler and can be rewritten:
\begin{equation}
\small
M_{rt} = \left(\begin{array}{cccc} 1 + i \epsilon_0 -\kappa
& \kappa  & \epsilon_0 & 0 \\
\kappa  & 1 - i \epsilon_0 - \kappa & 0 & -
\epsilon_0\\
- \epsilon_0 & 0 & 1 + i \epsilon_0-\kappa & \kappa  \\
0 & \epsilon_0 & \kappa & 1-i \epsilon_0- \kappa
\end{array}\right).
\end{equation}

The eigenvector of the round-trip equation, $M_{rt} \vec{\mathcal
J}= \vec {\mathcal J}$, is given by
\begin{equation}
\vec {\mathcal J} = \mathcal J \left(\begin{array}{c} 1\\1\\-i \\
-i \end{array}\right) \label{eq:solstatchamps}
\end{equation}
where $\mathcal J$ is a (\emph{a priori} complex) constant. This
expression shows that the field generated consists in two modes at
frequencies $\omega_a$ and $\omega_b$ with right circularly
polarization. $\mathcal J$ is determined by the pump depletion
equation:
\begin{eqnarray}
a_0 = \left(a_0^{in} - g a_1 b_2 \right) e^{i k n_0 l}
\end{eqnarray}
where $a_0^{in}$ is the input pump amplitude. In order to solve
this equation, one can set the pump phase between the two crystals
to be zero : this amounts to changing the phase origin which will
not change the properties of the system. The previous equation
yields
\begin{eqnarray}
\left(\frac{\kappa}{g} + g |\mathcal{J}|^2\right)^2=I_0^{(in)}
\end{eqnarray}
with $I_0^{(in)} = \left|a_0^{(in)}\right|^2$. One then gets the
usual OPO solution
\begin{equation}
\mathcal{J} = \sqrt{\frac{\kappa}{g^2} (\sigma -1)} e^{i \phi_1}
\label{eq:solstatv}
\end{equation}
where a reduced pumping parameter $\sigma$ has been defined equal
to the input pump amplitude normalized to the threshold
\begin{eqnarray}
\sigma  = \sqrt{\frac{I_0^{(in)}}{I_0 ^{(threshold)}}}.
\end{eqnarray}

When $\sigma$ is larger than 1, i.e. when the OPO is pumped above
a defined threshold, four bright beams with fixed relative phases
given by Eq. \ref{eq:solstatchamps} are generated. Since the two
signal (resp. idler) beams have a fixed phase relation and
identical frequencies, they form a single beam with a circular
polarization state. The system thus generates two circularly
polarized beams which are frequency tunable.

The next sections are devoted to the study of the quantum
properties. Correlations and anticorrelations of the couple
signal/idler emitted by each crystal, i.e. $(a_1,b_2)$ or
$(a_2,b_1)$, are formulated.

\section{Quantum properties and quadrature fluctuations}
\label{sec3}

In order to calculate the fluctuations for the involved fields
when the system is pumped above threshold, we apply the usual
input-output linearization technique (see
\emph{e.g.}~\cite{fabre90}). Individual noise spectra as well as
the correlations between any linear combinations of the various
fluctuations are derived in the Fourier domain.

\subsection{Linearized equations}

We first determine the classical stationary solutions of the
evolution equations : these solutions are denoted
$a_i^{(s)},b_j^{(s)}$ where $i=0,1,2;\, j=1,2$. We then linearize
the evolution equations around these stationary values by
setting~: $a_i = a_i^{(s)} + \delta a_i, \,i=0,\, 1,\, 2$ and $b_j
= b_j^{(s)} + \delta b_j, \,j=0,\, 1$.

The evolution equations for the fluctuations are deduced from the
round-trip matrix taking into account the stationary solutions
\ref{eq:solstatpompe},\ref{eq:solstatchamps},\ref{eq:solstatv}:
\renewcommand{\arraystretch}{2.5}
\begin{widetext}
\begin{equation}
\begin{array}{>{\ds}c>{\ds}c>{\ds}c>{\ds}c>{\ds}c>{\ds}c>{\ds}c>{\ds}c>{\ds}c>{\ds}c>{\ds}c>{\ds}c>{\ds}c}
\tau \frac{d \delta a_1}{dt} &=& (- \kappa \sigma + i \epsilon_0)
\delta a_1& + &\kappa \delta b_2^\ast &+ & e^{i \phi_1}
\sqrt{\kappa (\sigma-1)} \left(\delta a_0^{(x)}\right)^{(in)} &+
&\epsilon_0
\delta a_2 & - &e^{2 i \phi_1}\kappa (\sigma-1) \delta b_2-\sqrt{2\kappa}\delta a_1^{in}\\
\tau \frac{d \delta a_2}{dt} &=& (- \kappa \sigma + i \epsilon_0)
\delta a_2& + &\kappa \delta b_1^\ast &- & i e^{i \phi_1}
\sqrt{\kappa (\sigma-1)} \left(\delta a_0^{(y)}\right)^{(in)} & -
&\epsilon_0
\delta a_1 & + &e^{2 i \phi_1}\kappa (\sigma-1) \delta b_1-\sqrt{2\kappa}\delta a_2^{in}\\
\tau \frac{d \delta b_1}{dt} &=& (- \kappa \sigma + i \epsilon_0)
\delta b_1& + &\kappa \delta a_2^\ast & + & i e^{- i \phi_1}
\sqrt{\kappa (\sigma-1)} \left(\delta a_0^{(y)}\right)^{(in)} & +
&\epsilon_0
\delta b_2 & + &e^{-2 i \phi_1}\kappa (\sigma-1) \delta a_2-\sqrt{2\kappa}\delta b_1^{in}\\
\tau \frac{d \delta b_2}{dt} &=& (- \kappa \sigma + i \epsilon_0)
\delta b_2& + &\kappa \delta a_1^\ast & - & e^{-i \phi_1}
\sqrt{\kappa (\sigma-1)} \left(\delta a_0^{(x)}\right)^{(in)} & -
&\epsilon_0 \delta b_1 & - &e^{-2 i \phi_1}\kappa (\sigma-1)
\delta a_1-\sqrt{2\kappa}\delta b_2^{in}
\end{array}
\end{equation}
\end{widetext}
where $\tau$ stands for the cavity round-trip time. $\delta
a_i^{in}$ correspond to the vacuum fluctuations entering the
cavity through the coupling mirror.

\subsection{Quadrature fluctuations}

We define the amplitude and phase quadratures of a mode $a$ with
mean phase $\phi_a$ by respectively
\begin{eqnarray}
p_a &=& \delta a e^{-i \phi_a} + \delta a^\ast e^{i \phi_a} \nonumber \\
q_a &=& -i \left(\delta a e^{-i \phi_a} - \delta a^\ast e^{i
\phi_a}\right).
\end{eqnarray}
The evolution equations can be then expressed in matrix form~:
\begin{equation}
\tau \frac{d }{dt} \delta \mathcal{J} = M \delta \mathcal{J} +
\sqrt{\kappa(\sigma-1)} \delta
\mathcal{J}_{0}^{in}+\sqrt{2\kappa}\delta \mathcal{J}^{in}
\end{equation}
where
\begin{widetext}
\begin{eqnarray}
\small
M = \left(\renewcommand{\arraystretch}{1}
\begin{array}{cccc|cccc}
 - \kappa \sigma & - \epsilon_0 & 0 & \epsilon_0 & 0 &0  & -
 \kappa(\sigma-2) & 0  \\
\epsilon_0 & - \kappa \sigma & - \epsilon_0 &0  &0 &0 & 0& -
\kappa \sigma \\
0 & \epsilon_0 & - \kappa \sigma & - \epsilon_0 & -
\kappa(\sigma-2)  & 0 & 0 &0 \\
-\epsilon_0 & 0 & \epsilon_0& - \kappa \sigma & 0 &  - \kappa
\sigma & 0 &0\\
\hline 0 & 0  & - \kappa(\sigma-2) & 0&  - \kappa \sigma & -
\epsilon_0 & 0 & \epsilon_0\\
0 & 0 & 0 & - \kappa \sigma & \epsilon_0 & - \kappa \sigma  & -
\epsilon_0 & 0 \\
- \kappa(\sigma-2) & 0 & 0 & 0 & 0 & \epsilon_0 & - \kappa \sigma
 & - \epsilon_0 \\
0 & - \kappa \sigma & 0 & 0 & - \epsilon_0 & 0 & \epsilon_0 & -
\kappa \sigma
\end{array}
\right),
\end{eqnarray}
\end{widetext}
$ \delta \mathcal{J}$ is the column vector of the quadrature
components
\begin{eqnarray}\label{1}
\delta \mathcal{J} = ( p_{a_1},  q_{a_1},  p_{a_2},  q_{a_2},
p_{b_1},
 q_{b_1},  p_{b_2},  q_{b_2} ),
\end{eqnarray}
$\delta \mathcal{J}_{0}^{in}$ for the input pump fluctuations,
\begin{eqnarray}\label{2}
\delta \mathcal{J}_{0}^{in} &=& ( p_0^{(x)^{in}}, q_0^{(x)^{in}},
p_0^{(y)^{in}}, q_0^{(y)^{in}},\nonumber\\&&\qquad p_0^{(y)^{in}},
q_0^{(y)^{in}}, p_0^{(x^{in})}, q_0^{(x)^{in}}).
\end{eqnarray}
and $\delta {\mathcal{J}}^{in}$ the for the vacuum fluctuations
entering the cavity through the coupling mirror
\begin{eqnarray}\label{3}
\delta {\mathcal{J}}^{in} &=& ( p_{a_1}^{in}, q_{a_1}^{in},
p_{a_2}^{in}, q_{a_2}^{in} ,\nonumber\\&&\qquad  p_{b_1}^{in},
q_{b_1}^{in}, p_{b_2}^{in},q_{b_2}^{in}).
\end{eqnarray}

These differential equations are readily transformed into
algebraic equations by taking the Fourier transform. In the
frequency domain, the equations become
\begin{equation}
2 i \Omega \delta\tilde{\mathcal{J}} (\Omega) = M' \delta
\tilde{\mathcal{J}} (\Omega) + \sqrt{\frac{\sigma-1}{\kappa}}
\delta \tilde{\mathcal{J}_{0}^{in}} (\Omega) + \sqrt{\frac 2
\kappa} \delta \tilde{\mathcal{J}}^{in} (\Omega).
\end{equation}
where $\Omega = \frac{\tau \omega}{2\kappa} = \frac{ \omega }{
\Omega_c }$ is the noise frequency normalized to the cavity
bandwidth $\Omega_c=\frac{2\kappa}{\tau}$ and
$c=\frac{\epsilon_0}{\kappa}$ is the normalized coupling constant.
$ \delta \tilde{\mathcal{J}} (\Omega)$, $\delta
\tilde{\mathcal{J}}_{0}^{in} (\Omega)$ and $\delta
\tilde{\mathcal{J}}^{in} (\Omega)$ are the Fourier transforms of
the column vectors \ref{1}, \ref{2}, \ref{3}. The matrix $M'$ is
defined by
\begin{eqnarray}
\small
M' = \left(\renewcommand{\arraystretch}{1}
\begin{array}{cccc|cccc}
 - \sigma & - c & 0 & c & 0 & 0  & - \sigma-2 & 0  \\
c & - \sigma & - c &0  &0 &0 & 0& - \sigma \\
0 & c & - \sigma & - c & -\sigma-2 & 0 & 0 &0 \\
-c & 0 & c& - \sigma & 0 &  - \sigma & 0 &0\\
\hline
0 & 0 & - \sigma-2 & 0&  - \sigma & - c & 0 & c\\
0 & 0 & 0 & - \sigma & c & - \sigma  & - c & 0 \\
- \sigma-2 & 0 & 0 & 0 & 0 & c & - \sigma & - c \\
0 & - \sigma & 0 & 0 & - c & 0 & c & - \sigma
\end{array}
\right).
\end{eqnarray}

\subsection{Correlations and anticorrelations}

In order to use the symmetry of the equations, one can introduce
the symmetric and antisymmetric components
\begin{eqnarray}
\tilde p_\alpha &=& \frac{1}{\sqrt 2} \left(\tilde p_{a_1}(\Omega)
+ \tilde p_{b_2}(\Omega)\right)\nonumber \\ \tilde q_\alpha &=&
\frac{1}{\sqrt
2} \left(\tilde q_{a_1}(\Omega) + \tilde q_{b_2} (\Omega) \right)\nonumber \\
\tilde r_\alpha &=& \frac{1}{\sqrt 2} \left(\tilde p_{a_1}
(\Omega) -\tilde p_{b_2} (\Omega) \right)\nonumber\\ \tilde
s_\alpha &=& \frac{1}{\sqrt 2} \left(\tilde q_{a_1} (\Omega)
-\tilde q_{b_2} (\Omega) \right)
\end{eqnarray}
and
\begin{eqnarray}
\tilde p_\beta &=& \frac{1}{\sqrt 2} \left(\tilde p_{a_2}(\Omega)
+ \tilde p_{b_1}(\Omega)\right)\nonumber\\ \tilde q_\beta &=&
\frac{1}{\sqrt 2} \left(\tilde q_{a_2}(\Omega) + \tilde q_{b_1}
(\Omega) \right)\nonumber\\
 \tilde r_\beta &=& - \frac{1}{\sqrt 2} \left(\tilde p_{a_2} (\Omega)
-\tilde p_{b_1} (\Omega) \right)\nonumber\\ \tilde s_\beta &=& -
\frac{1}{\sqrt 2} \left(\tilde q_{a_2} (\Omega) -\tilde q_{b_1}
(\Omega) \right).
\end{eqnarray}
In a standard OPO, one expects intensity correlations and phase
anti-correlations. Using our notations, this corresponds to having
$r_\alpha$ and $r_\beta$ as well as $q_\alpha$ and $q_\beta$
squeezed \cite{reynaud87}. This is also the case in a phase-locked
OPO in the limit of small coupling \cite{epjdquantique}.

As previously, the equations verified by these quantities can be
expressed in matrix form. However, it is interesting to note that
the sums are independent from the differences. Thus, one can write
two sets of equations:
\begin{eqnarray}
2 i \Omega \delta\tilde w_\pm (\Omega) &=& M_\pm \delta \tilde
w_\pm (\Omega) \\&&+ \sqrt{2\frac{\sigma-1}{\kappa}} \delta \tilde
w_{0,\pm}^{in} (\Omega) + \sqrt{\frac 2 \kappa} \delta \tilde
w_\pm^{in} (\Omega)\nonumber
\end{eqnarray}
where $ \delta \tilde w_\pm$ are the column vector
\begin{eqnarray}
\delta\tilde w_+ (\Omega)&=& \left(\tilde p_\alpha (\Omega),
\tilde p_\beta (\Omega), \tilde q_\alpha (\Omega), \tilde q_\beta
(\Omega)\right),\nonumber\\\delta\tilde w_- (\Omega)&=& \left(
\tilde r_\alpha (\Omega), \tilde r_\beta (\Omega), \tilde s_\alpha
(\Omega), \tilde s_\beta (\Omega) \right),
\end{eqnarray}
\begin{eqnarray}
\small M_+ &=& \left(\renewcommand{\arraystretch}{1.5}
\begin{array}{cccc}
 - 2 (\sigma-1) & 0 & - c & c   \\
0 & -2( \sigma-1) &  c &-c  \\
c & -c & - 2 \sigma & 0  \\
-c & c & 0& -2 \sigma
\end{array}
\right), \nonumber \\ M_- &=&
\left(\renewcommand{\arraystretch}{1.5}\begin{array}{cccc}
  - 2 &  0 & - c & -c\\
 0 & - 2  &  -c & -c \\
  c &  c & 0 & 0\\
 c &  c & 0 & 0
\end{array}
\right),
\end{eqnarray}
$\delta \tilde w_{0,\pm}^{in}$ are the column vectors for the
input pump fluctuations,
\begin{eqnarray}
\delta \tilde w_{0,+}^{in} &=& \left( \tilde p_0^{(x)^{in}}
(\Omega), \tilde p_0^{(y)^{in}} (\Omega), \tilde q_0^{(x)^{in}}
(\Omega), \tilde p_0^{(y)^{in}} (\Omega)\right)\nonumber\\\delta
\tilde w_{0,-}^{in} &=& \left( 0, 0, 0, 0 \right)
\end{eqnarray}
and $\delta \tilde w^{in}$ the column vector for the input
fluctuations entering the cavity through the coupling mirror
\begin{eqnarray}
\delta\tilde w_+^{in} (\Omega)&=& \left(\tilde p_{\alpha}^{in}
(\Omega), \tilde p_\beta^{in} (\Omega), \tilde q_{\alpha}^{in}
(\Omega), \tilde q_\beta^{in} (\Omega)\right) \nonumber \\
\delta\tilde w_-^{in} (\Omega)&=& \left( \tilde r_\alpha^{in}
(\Omega), \tilde r_\beta^{in} (\Omega), \tilde s_\alpha^{in}
(\Omega), \tilde s_\beta^{in} (\Omega) \right)
\end{eqnarray}

\begin{figure*}[htpb!]
\centerline{\includegraphics[width=.9\columnwidth]{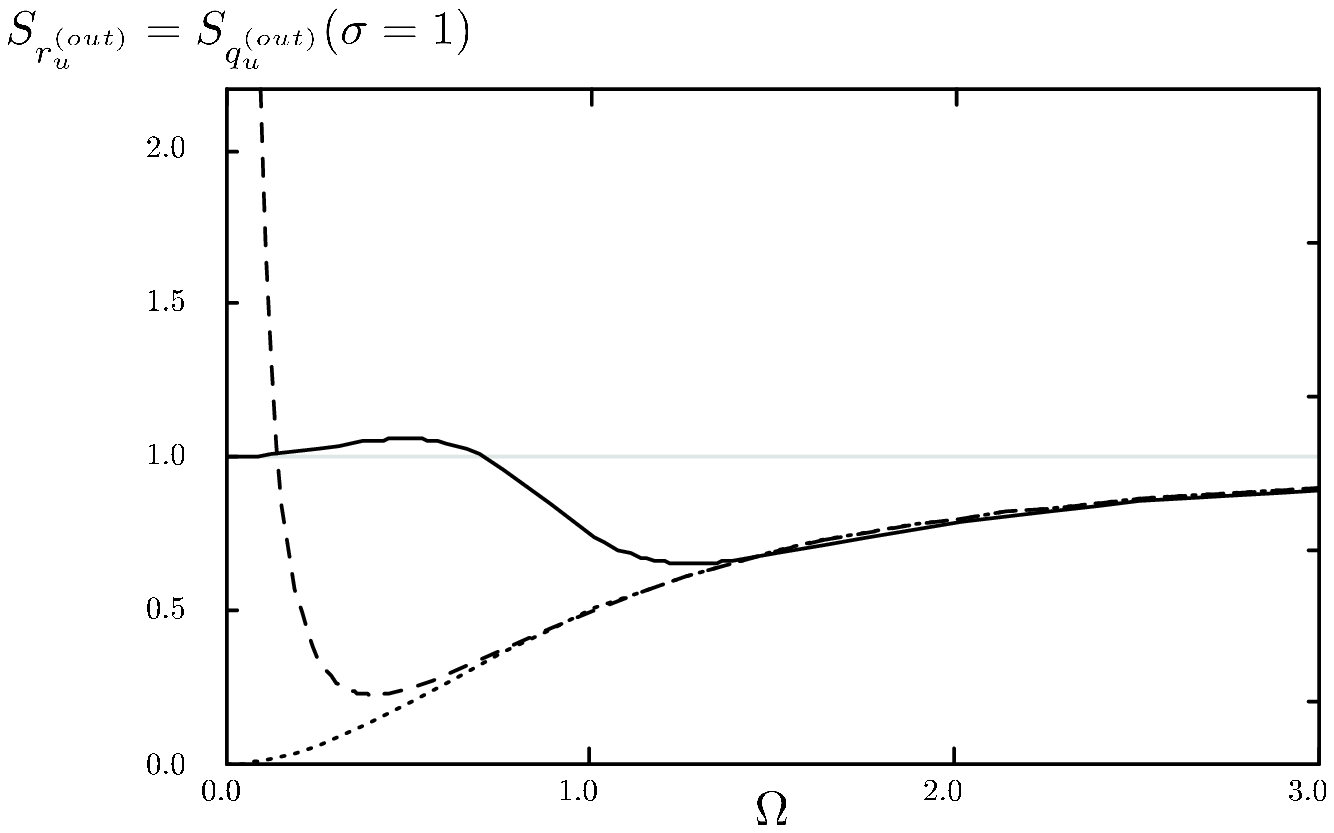}%
\includegraphics[width=.9\columnwidth]{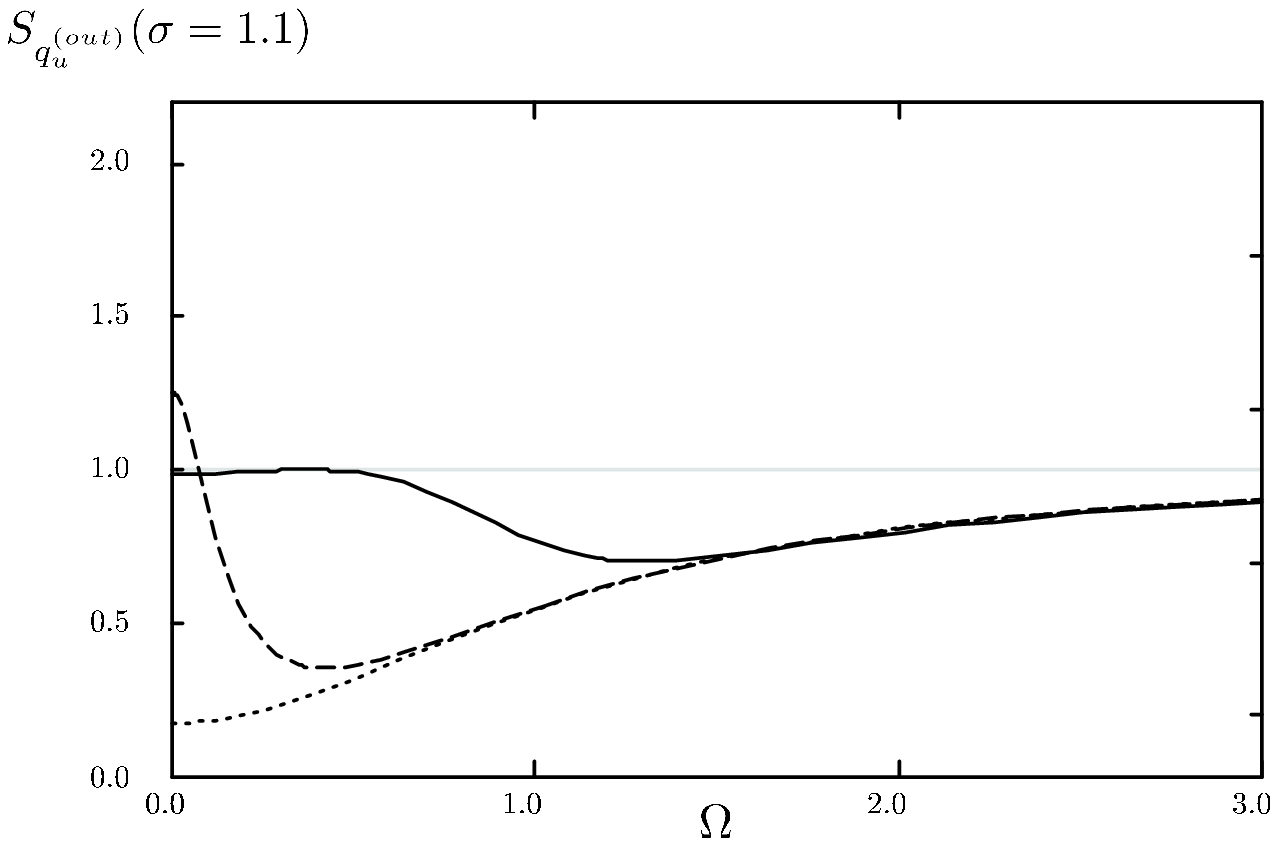}}
\caption{\label{fig:spectre_rq}Normalized noise spectrum of the
amplitude quadrature difference and phase quadrature sum for
$\sigma = 1$ (left) and $\sigma = 1.1$ (right). Continuous curves
corresponds to $c=1$, long-dashed to $c=0.2$ and short-dashed to
$c=0$.}
\end{figure*}

These equations can be easily solved and one obtains the equations
for the intracavity fluctuations. However, the relevant
fluctuations are those outside the cavity. They can be easily
calculated using the boundary condition on the output mirror:
\begin{eqnarray}
f_{out} = t f - r f_{in} \approx \sqrt{2\kappa} f - f_{in}
\end{eqnarray}
and the corresponding spectra are given by
\begin{eqnarray}
\EuScript{S}_f (\Omega )= \left\langle \tilde f_{out} (\Omega)
\tilde f_{out}^\ast (-\Omega)\right\rangle.
\end{eqnarray}
The expressions are identical for the two sets of spectra
$\EuScript{S}_{p_\alpha,q_\alpha,r_\alpha,s_\alpha}$ and
$\EuScript{S}_{p_\beta,q_\beta,r_\beta,s_\beta}$:
\begin{widetext}
\begin{eqnarray}
\EuScript{S}_{p_u}  (\Omega )&=& 1 + \frac{1}{2(\Omega^2 + (\sigma
-1)^2}) + \frac{\sigma^2 + \Omega^2 - c^2}{2\left((c^2 + \sigma
(\sigma -1))^2 + \Omega^2  - 2 (c^2 -\sigma(\sigma-1))\Omega^2 +
\Omega^4\right)},\nonumber\\
\EuScript{S}_{q_u}  (\Omega )&=& 1 - \frac{1}{\Omega^2 + \sigma^2}
- \frac{(\sigma-1)^2 + \Omega^2 - c^2}{2\left((c^2 + \sigma
(\sigma -1))^2 + \Omega^2 - 2 (c^2 -\sigma(\sigma-1))\Omega^2 +
\Omega^4\right)}, \nonumber\\
\EuScript{S}_{r_u}  (\Omega )&=& 1 - \frac{1}{2(1+ \Omega^2)} -
\frac{\Omega^2 -
c^2}{2\left(\Omega^2 + (\Omega^2-c^2)^2\right)},\nonumber\\
\EuScript{S}_{s_u}  (\Omega ) &=& 1 + \frac{1}{2 \Omega^2} +
\frac{1 + \Omega^2 - c^2}{2\left(\Omega^2 +
(\Omega^2-c^2)^2\right)},
\end{eqnarray}
$u=\alpha,\beta$.
\end{widetext}
The behavior of the squeezed spectra, $\EuScript{S}_{r_u^{(out)}}$
and $\EuScript{S}_{q_{u}^{(out)}}$, is shown on figure
\ref{fig:spectre_rq}. For small coupling and low analysis
frequency, one observes that the spectra go to zero which
indicates the existence of very large correlations. As usual, the
spectra go to one as $\Omega$ goes to infinity: the correlations
exist only within the cavity bandwidth or conversely, only when
the integration time is larger than the field life time in the
cavity. The increase of $c$ also results in a degradation of the
correlations and anti-correlations: the wave-plate tends to mix
the fields, and this phenomenon is increased as the coupling is
increased. This phenomenon can be understood in terms of phase
diffusion: as the coupling is increased, the phase diffusion is
reduced which in turn reduces the intensity correlations. Finally,
the phase anticorrelations (Fig. \ref{fig:spectre_rq}, left) are
also degraded as the input pump power is increased : as $\sigma$
is increased, the input pump noise becomes more and more
important.

This analysis has been performed by studying the fields generated
by each crystal. Since several frequencies are involved, such a
measurement would require the use of multiple local oscillators or
of dephasing cavities \cite{saopaulo}. It is thus easier
experimentally as well as more fruitful for quantum information
purposes to study the beams at each frequency, namely the couples
$(a_1,a_2)$ and $(b_1,b_2)$, which are circularly polarized.

\section{Polarization entanglement generation}
\label{sec4}

The non-classical properties of the two-crystal OPO are now
analyzed in terms of polarization fluctuations. Polarization
entanglement generation is demonstrated using various criteria
introduced in the literature.

\subsection{Quantum Stokes parameters}

Before describing the properties of the system, let us briefly
recall the quantum picture for polarization properties. The
polarization of light beams can be described in the general case
by the Stokes parameters \cite{stokes} :
\begin{eqnarray}
S_0 &=& I_x + I_y \nonumber\\
S_1&=& I_x - I_y \nonumber\\
S_2 &=& I_{+45^\circ} - I_{-45^\circ}\nonumber\\
S_3 &=& I_{\sigma^+} - I_{\sigma^-}.
\end{eqnarray}
where $I_{x,y,+45^\circ,-45^\circ,\sigma^+,\sigma^-}$ describe the
intensity along the $x$, $y$, $+45^\circ$, $-45^\circ$,
$\sigma^+$, and $\sigma^-$ polarization. These four parameters are
not independent but are instead linked by the relation
\begin{eqnarray} S_0^2 = S_1^2 + S_2^2 + S_3^2
\end{eqnarray} for a totally polarized field.

The quantum polarization properties are defined via the quantum
counter-parts of the Stokes parameters
\cite{chirkin93,korolkova02}:
\begin{eqnarray}
\hat S_0 &=& \hat a^\dagger_x \hat a_x + \hat a^\dagger_y \hat
a_y\nonumber\\
\hat S_1 &=& \hat a^\dagger_x \hat a_x - \hat a^\dagger_y \hat
a_y\nonumber\\
\hat S_2 &=& \hat a^\dagger_x \hat a_y + \hat a^\dagger_y \hat
a_x= \hat a^\dagger_{+45^\circ} \hat a_{+45^\circ} + \hat
a^\dagger_{- 45^\circ} \hat a_{- 45^\circ}\nonumber\\
\hat S_3 &=& i \left(\hat a^\dagger_y \hat a_x - \hat a^\dagger_x
\hat a_y \right) = \hat a^\dagger_{\sigma^+} \hat a_{\sigma^+} +
\hat a^\dagger_{\sigma^-} \hat a_{\sigma^-}
\end{eqnarray}
$(\hat S_1, \hat S_2, \hat S_3)$ verify commutation relations
identical to those verified by orbital momentum operators:
\begin{eqnarray}
[\hat S_i, \hat S_j ] = 2 i  \hat S_k
\end{eqnarray}
where $i,j,k=1,2,3$ are cyclically interchangeable. These
relations result in three uncertainty relations,
\begin{eqnarray}
\Delta^2 \hat S_1 \Delta^2 \hat S_2 &\geq& \left|\langle \hat S_3
\rangle \right|^2, \,\Delta^2 \hat S_2 \Delta^2 \hat S_3 \geq
\left|\langle \hat S_1 \rangle \right|^2, \nonumber\\\Delta^2 \hat
S_3 \Delta^2 \hat S_1 &\geq& \left|\langle \hat S_2 \rangle
\right|^2. \label{eq:heisenberg stokes}
\end{eqnarray}
where the notation $\Delta^2 (\hat X)$ correspond to the variance
of the operator $\hat X$.

\subsection{Entanglement criteria}

In our case, Eq. \ref{eq:solstatchamps} shows that the output
beams are circularly polarized, thus $\langle \hat S_1^{a,b}
\rangle = 0 = \langle \hat S_2^{a,b} \rangle$. It results that the
inequalities \ref{eq:heisenberg stokes} which contain variances
$\hat S_3^{a,b}$ are automatically verified. The only non-trivial
Heisenberg inequality is then given by
\begin{eqnarray}
\Delta^2 \hat S_1^{a,b} \Delta^2 \hat S_2^{a,b} \geq \left|\langle
\hat S_3^{a,b} \rangle \right|^2 = \frac{\kappa}{g^2} (\sigma-1)
\end{eqnarray}
where the notation $\Delta^2 (\hat X)$ correspond to the variance
of the operator $\hat X$: $\Delta^2 (\hat X) = \left\langle
\left(\hat X - \langle \hat X \rangle \right)^2\right\rangle =
\left\langle \delta X^2\right\rangle $.

Various entanglement criteria have been defined. We will restrict
ourselves to three of them \cite{duan00,simon00,mancini,reid}.

The first two criteria are, in the general case, sufficient (and
not always necessary) conditions for entanglement. A general
criterion is given by the product of two linear combinations of
conjugate variables variances \cite{mancini} (denoted "product
criterion" in the following). In the case of the Stokes operators,
this inseparability criterion states that when
\begin{equation}
\Delta^2 (\hat S_1^a \pm \hat S_1^b) . \Delta^2 (\hat S_2^a \mp
\hat S_2^b) \leq 2 \left(\left|\langle \hat S_3^{a} \rangle
\right| + \left|\langle \hat S_3^{b} \rangle \right|\right)
\label{eq:mancini}
\end{equation}
the state is entangled. The relevant Stokes parameters
fluctuations, $\delta \hat S_1^{a,b}$ and $\delta \hat S_2^{a,b}$,
can be expressed from the quadrature operators. In the case of
circularly polarized beams, one has ($u=a,b$)
\begin{eqnarray}
\frac{\delta \hat S_1^{u} (\Omega)}{|{\mathcal{J}}|} &= & \delta
\tilde p_{u_1}^{(out)} (\Omega) - \delta \tilde p_{u_2}^{(out)}
(\Omega),\nonumber\\ \frac{\delta \hat S_2^u
(\Omega)}{|{\mathcal{J}}|} &= & -(\delta \tilde q_{u_1}^{(out)}
(\Omega) - \delta \tilde q_{u_2}^{(out)} (\Omega)).
\end{eqnarray}
It follows from these expressions that:
\begin{widetext}
\begin{eqnarray}
\delta S_1 ^+ = \delta \hat S_1^a (\Omega) + \delta \hat S_1^b
(\Omega) = (\delta \tilde p_{a_1}^{(out)} (\Omega) - \delta \tilde
p_{b_2}^{(out)} (\Omega)) - (\delta \tilde p_{a_2}^{(out)}
(\Omega) - \delta \tilde p_{b_1}^{(out)} (\Omega)) = \sqrt 2 (
r_\alpha^{(out)} - r_\beta^{(out)}) \nonumber\\
\delta S_2 ^- = \delta \hat S_2^a (\Omega) - \delta \hat S_2^b
(\Omega) = -(\delta \tilde q_{a_1}^{(out)} (\Omega) + \delta
\tilde q_{b_2}^{(out)} (\Omega)) + (\delta \tilde q_{a_2}^{(out)}
(\Omega) + \delta \tilde p_{b_1}^{(out)} (\Omega)) = \sqrt 2
(-q_\alpha^{(out)} + q_\beta^{(out)})
\end{eqnarray}
\end{widetext}
Thus the relevant spectra can be calculated from the previously
calculated linear combinations of the amplitude and phase
quadrature components. They are given by:
\begin{widetext}
\begin{eqnarray}
\EuScript{S}_{S_1^+} (\Omega) &=& 1- \frac{\Omega^2-c^2}{\Omega^2
+
(\Omega^2-c^2 )^2} \label{eq:s1plus} \nonumber\\
\EuScript{S}_{S_2^-} (\Omega)  &=&1-
\frac{(\Omega^2-c^2)+(\sigma-1)^2}{\left[ c^2 + \sigma (\sigma
-1)\right]^2 + \Omega^2 - 2 \left[c^2 - \sigma (\sigma-1)\right]
\Omega^2 + \Omega^4}. \label{eq:s2moins}
\end{eqnarray}
\end{widetext}

\begin{figure*}[htpb!]
\centerline{\includegraphics[width=.8\columnwidth]{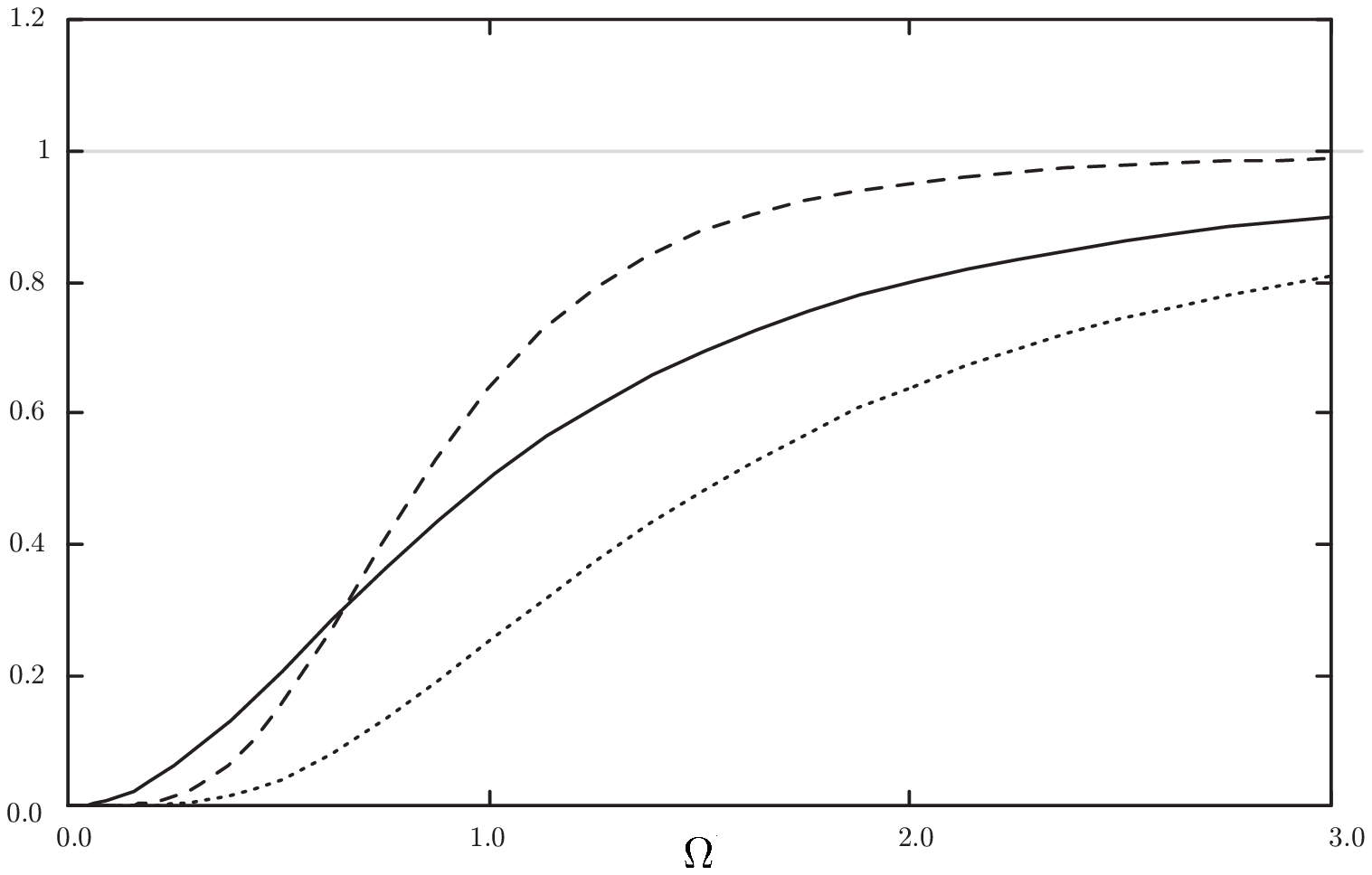}%
\includegraphics[width=.8\columnwidth]{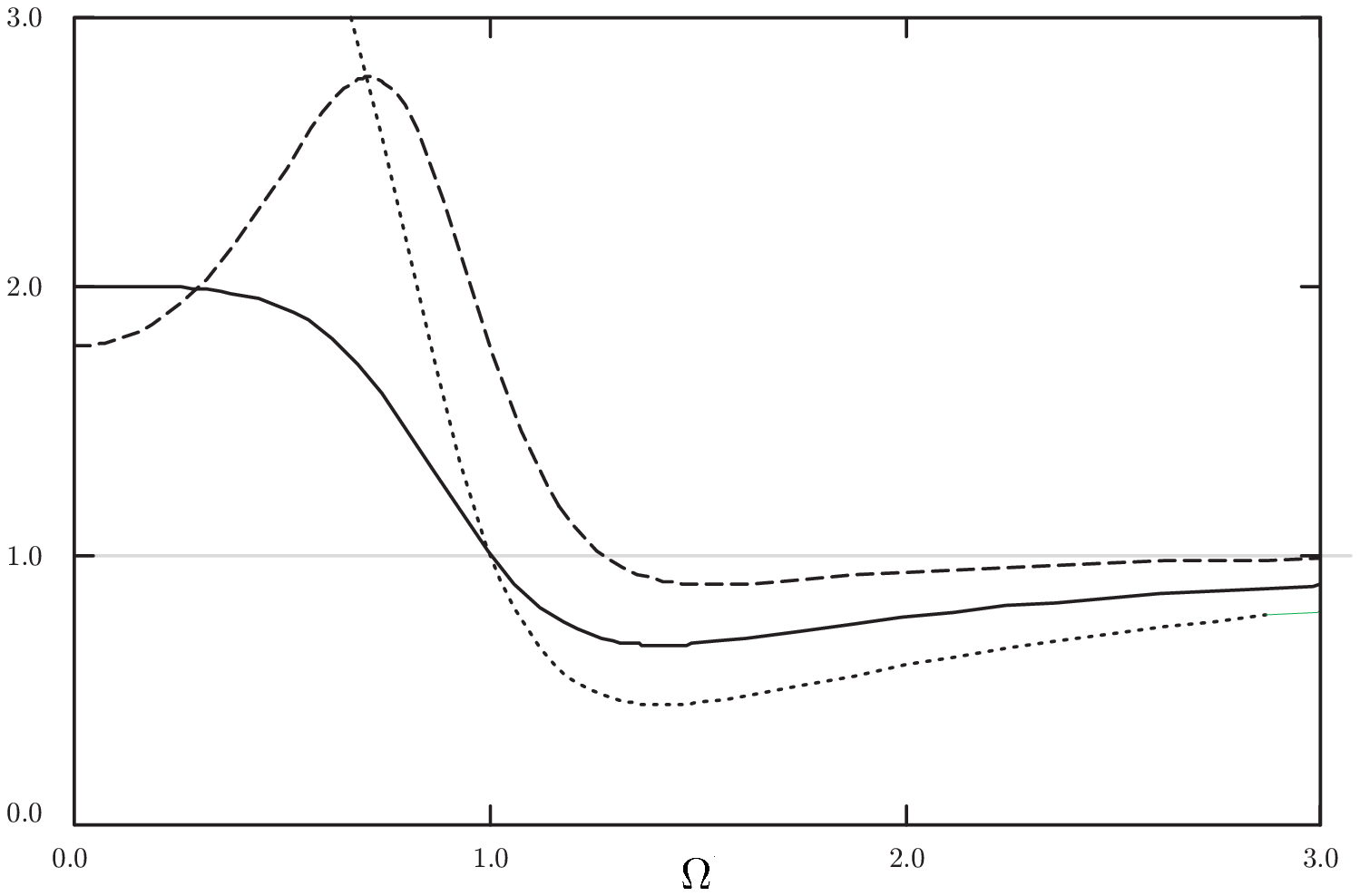}}
\caption{\label{fig:criteres}Criterion spectra for $c=0$ (left)
and $c = 1$ (right). Continuous curves correspond to the sum
criterion ($\frac{1}{2}\EuScript{S}_{S_1^+} (\Omega) +
\EuScript{S}_{S_2^-} (\Omega) $), long-dashed to the EPR criterion
($\EuScript{S}_{S_1^a |S_1^b} .  \EuScript{S}_{S_2^a |S_2^b}$) and
short-dashed to the product criterion
($\frac{1}{2}\left(\EuScript{S}_{S_1^+} .
\EuScript{S}_{S_2^-}\right)$).}
\end{figure*}
Let us remark that the two expressions are equal for $\sigma = 1$
and $c=0$, \emph{i.e.} for the uncoupled system operated at
threshold. One also remarks that $\EuScript{S}_{S_1^+} (\Omega) $
is independent of the pump amplitude, $\sigma$. This is not
surprising since this quantity is the difference of the intensity
correlations between the signal and idler modes of each crystal.
These correlations originates in the parametric down-conversion
phenomenon, independently of the pump power: the photons in the
modes $a_1$ and $b_2$ ($a_2$ and $b_1$ respectively) are emitted
by pairs thus leading to intensity correlations between the two
modes. These two quantities can be measured using standard methods
for the measurement of quantum Stokes operators, see \emph{e.g.}
\cite{korolkova02,polarentanglement_bowen,polarentanglement_schnabel}

Using these expressions, one gets a simple expression for the
product criterion \ref{eq:mancini}
\begin{equation}
\EuScript{S}_{S_1^+} (\Omega) \EuScript{S}_{S_2^-} (\Omega) \leq
2.
\end{equation}

The second criterion ("sum criterion"), which is widely used in
continuous-variable entanglement characterization, is given by the
sum of the above quantities \cite{duan00,simon00}. It is in fact a
special case of the previous criterion \cite{mancini}. The
criterion states that if
\begin{equation}
\frac{1}{2}\left(\EuScript{S}_{S_1^+} +
\EuScript{S}_{S_2^-}\right) \leq 1
\end{equation}
then the two states are entangled.

These two criteria are entanglement witnesses in the sense that
they allow to specify whether the state is indeed entangled.

Finally, the last criterion is related to another particularity of
the system which may allow to make a joint QND measurement of the
two non-commutating observables $\hat S_1$ and $\hat S_2$. It is
shown in \cite{reid} that this is related to EPR-like "paradox".
This criterion relies on conditional probabilities and can be
written
\begin{equation}
\EuScript{S}_{S_1^a |S_1^b} .  \EuScript{S}_{S_2^a |S_2^b} \leq 1,
\end{equation}
where \begin{eqnarray} \EuScript{S}_{S_i^a |S_i^b} = \langle
(S_i^a)^2 \rangle \left(1 - \frac{\langle S_i^a S_i^b
\rangle^2}{\langle (S_i^a)^2 \rangle\langle (S_i^b)^2
\rangle}\right) , \quad i=1,2.
\end{eqnarray}
When this condition is verified, the correlations are sufficiently
large that the information extracted from the measurement of the
two Stokes operators of one field provides values for the Stokes
operators of the other which violate the Heisenberg inequality
\cite{treps04}.

Figure \ref{fig:criteres} shows the various criteria as a function
of the analysis frequency, $\Omega$ for a fixed pump power
($\sigma=1$) and for different values of the coupling. Very strong
entanglement can be found for small couplings, $c$ and small
analysis frequencies, $\Omega$. As the coupling is increased, the
entanglement is shifted to higher values of the analysis
frequency. Let us note however that $c=1$ corresponds usually to
large angles (around 1$^\circ$ for typical output couplers): in
the case of single crystal phase locked OPOs, stable phase-locking
has been achieved with very small angles corresponding to
$c\approx 0$. The entanglement is present for a wide range of the
parameters thus showing the efficiency of the system.

\section{Conclusion}

We have presented and studied theoretically an original device
based on an optical cavity containing two type-II phase-matched
$\chi^{(2)}$ crystals as well as birefringent plates which add a
linear coupling between the signal and idler fields emitted in
each crystal. The coupling induces a phase-locking respectively
between the two signal fields and between the two idler fields. In
this configuration, the system is found to directly generate
two-color polarization-entangled beams without the need for mode
mixing. The two-crystal OPO would be thus a useful resource for
the field of continuous-variable quantum information.
Non-classical polarization states are well coupled to atomic
ensembles which can be used as quantum memories and form the basic
block required for quantum networks. Furthermore, we believe that
the experimental implementation of this device is very interesting
for potential parallel processing as it would be a new step
towards extending the dimension of the phase space experimentally
accessible.

\section*{Acknowledgements}

Laboratoire Kastler-Brossel, of the Ecole Normale Sup\'{e}rieure
and the Universit\'{e} Pierre et Marie Curie, is associated with
the Centre National de la Recherche Scientifique (UMR 8552).
Laboratoire Mat{\'e}riaux et Ph{\'e}nom{\`e}nes Quantiques, of
Universit\'e Denis Diderot, is associated with the Centre National
de la Recherche Scientifique (UMR 7162).

We thank A. Gatti for pointing us to ref. \cite{mancini} and
acknowledge fruitful discussions with G. Adesso and A. Serafini.

\end{document}